\documentclass[twocolumn,aps,prl,amsmath,amssymb,floatfix]{revtex4-2}
\usepackage{graphicx}
\usepackage{dcolumn}
\usepackage{bm}
\usepackage{xcolor}
\usepackage[normalem]{ulem}

\begin{document}

\title{The Minimum Leidenfrost Temperature on Smooth Surfaces}

\author{Dana Harvey}
\author{Joshua M\'endez Harper}
\author{Justin C. Burton}
\email{justin.c.burton@emory.edu}

\affiliation{Department of Physics, Emory University, Atlanta, Georgia 30322, USA}

\date{\today}

\begin{abstract}
During the Leidenfrost effect, a thin insulating vapor layer separates an evaporating liquid from a hot solid. Here we demonstrate that Leidenfrost vapor layers can be sustained at much lower temperatures than those required for formation. Using a high-speed electrical technique to measure the thickness of water vapor layers over smooth, metallic surfaces, we find that the explosive failure point is nearly independent of material and fluid properties, suggesting a purely hydrodynamic mechanism determines this threshold. For water vapor layers of several millimeters in size, the minimum temperature for stability is $\approx$ 140$^\circ$C, corresponding to an average vapor layer thickness of 10-20 $\mu$m. 
\end{abstract}

\maketitle

In his seminal 1756 treatise, J. G. Leidenfrost noted that a water droplet placed on a heated, polished metal spoon does not wet the surface \cite{Leidenfrost}. Instead, the water droplet levitates above the hot surface, cushioned by a vapor film generated by evaporation. Since then, the Leidenfrost effect has been well studied due to its importance in laboratory, industrial, and geophysical contexts \cite{Querer}. Examples include the vapor layer geometry \cite{Burton,Sobac2014,Biance,Snoeijer2009,duchemin2005static,lister2008shape}, spontaneous motion and oscillations of drops \cite{Bouillant2,Cousins2012,Lagubeau2011,Linke2006,Gauthier2019,Ma2,Ma}, drop impact on heated surfaces \cite{tran2012drop,castanet2015drop,riboux_gordillo_2016,YAO1988363}, and ``nano-painting'' through particle deposition \cite{Bain,Elbahri}. 

In nature, the Leidenfrost effect--or more precisely, the collapse of a Leidenfrost vapor layer between ascending magma and an aquifer--underpins one of the most energetic and common forms of volcanism: phreatomagmatic eruptions \cite{lorenz2003maar, de2020stratigraphy}. The Leidenfrost effect need not involve water or even a liquid; blocks of sublimating CO$_2$ ice may ``surf'' down Martian dunes on lubricating layers of CO$_2$ gas, carving channels and pits on the red planet's surface \cite{mc2017experiments}. This very manifestation of the Leidenfrost effect may help power the first Martian colonies \cite{wells2015sublimation}. 

In all of these examples, precise knowledge of the transition temperature at which the vapor layer forms (or fails) is crucial. However, reported values of this temperature vary widely in the literature, and are known to depend on surface roughness \cite{Kim2,Kim,Kruse,KimNano,Jones2019}, hydrophobicity \cite{Liu,Vakarelski,Vakarelski2,KimNano}, thermal properties of the solid \cite{Freud,Vakarelski, Vakarelski2,Yagov,Hsu,Sher}, liquid temperature \cite{Jouhara,Freud,Sher,Yagov,Yagov2,KimNano}, solid geometry \cite{Bradfield,Huang,Jouhara,Freud, Vakarelski,Vakarelski2,Sher,Yagov}, and liquid impurities \cite{Huang,Abdalrahman,Hsu,KimNano}. For smooth, homogeneous surfaces, a comprehensive theoretical study by Zhao et al. \cite{Zhao} showed that the temperature at which the vapor layer forms spontaneously from a liquid-solid contact depends only on the hydrophobicity of the surface \cite{Zhao}. For water drops on metallic surfaces, this corresponds to temperatures exceeding 200$^\circ$C. Yet, once formed, Leidenfrost drops can exist on metal surfaces with temperatures below boiling temperature (100$^\circ$C) \cite{NASA}. 

Here, we show how this large metastable region between formation and failure arises from the hydrodynamic stability of the gas flow in the vapor layer. Our experiments employ a new electrical technique that can directly measure the average thickness of the vapor layer around a heated solid with microsecond resolution.  For smooth metallic surfaces, we find a formation Leidenfrost temperature, $T_+$, consistent with recent predictions of a nanoscale wetting theory \cite{Zhao}. Once a stable vapor layer is formed over a given solid surface, its thickness is solely a function of the surface temperature, $T_s$. Remarkably, we find a minimum Leidenfrost temperature, $T_-$, which is nearly independent of liquid impurities and solid properties. At this temperature, the vapor layer spontaneously fails through liquid-solid contact and rapid boiling.

\begin{figure}[!]
\centering
\includegraphics[width=3.25in]{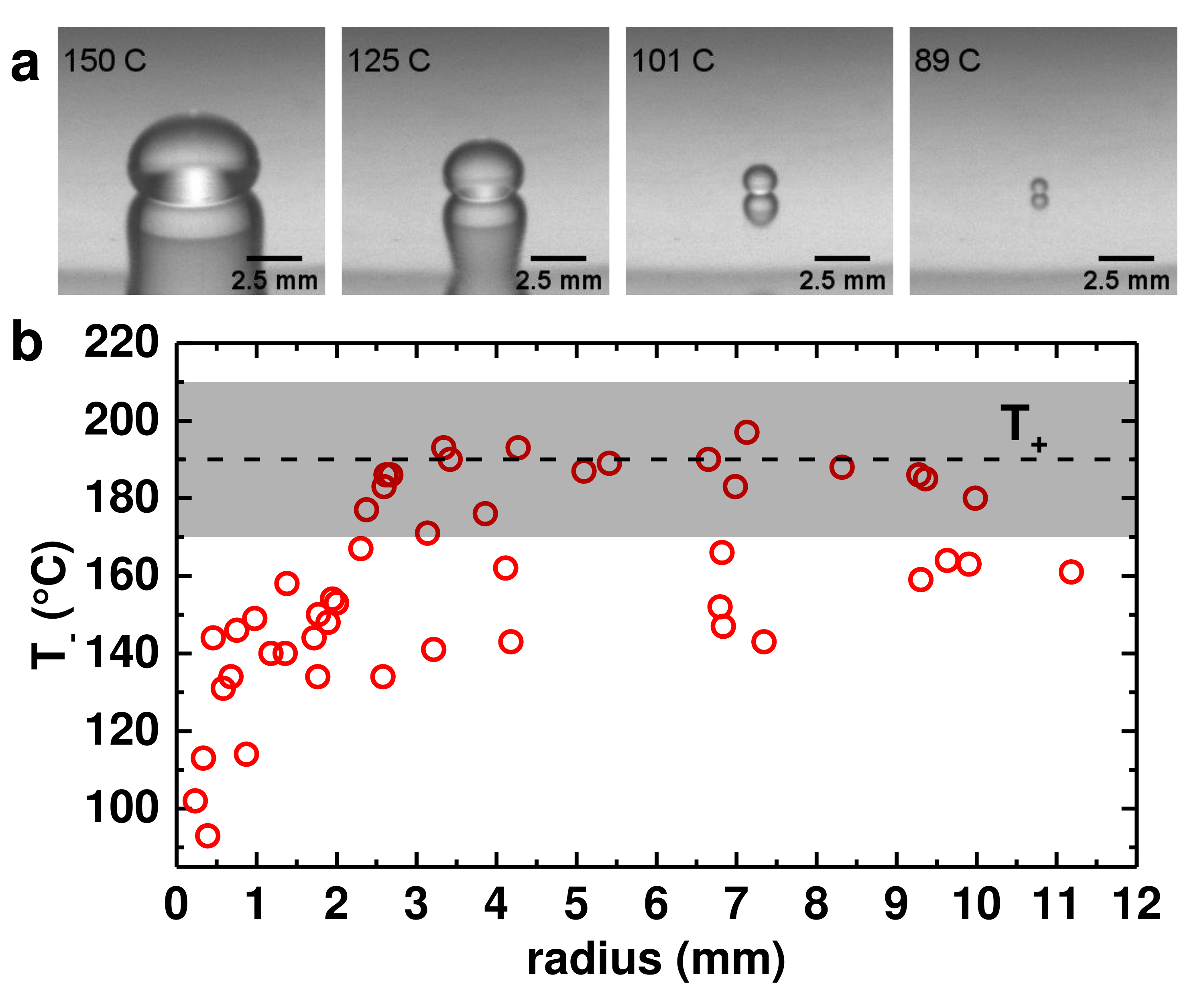}
\caption{(a) Sequence of images showing a single pure water Leidenfrost drop that remains levitated over a heated aluminum surface below the boiling point (Video S1 \cite{supp}). (b) Failure temperature, $T_-$, as a function of drop radius. Each data point represents a different water drop. The upper Leidenfrost temperature, $T_+ = 190 \pm 20^{\circ}$C, is indicated by the dashed line and shaded region.}
\label{Fig1}
\end{figure}

Figure \ref{Fig1}a illustrates how a small water droplet can levitate over a heated concave aluminum surface with temperature $T_s<T_+$. As the surface cools, the drop evaporates, but remains levitated (Video S1 \cite{supp}). Eventually, liquid-solid contact occurs at $T_{-}$. This minimum temperature varies with drop size (Fig.\ \ref{Fig1}b, Video S2 \cite{supp}). Surprisingly, droplets smaller than $\approx$ 100 $\mu$m can even levitate below the boiling point by way of a diffusive Stefan flow \cite{Zaitsev}. Conversely, droplets with radii larger than the capillary length of water, $l_c \sim $ 2.5 mm, fail at higher temperatures that are roughly independent of drop size. Large variations in $T_-$  exist in this regime, possibly due to vapor layer oscillations sustained by evaporation \cite{Ma2,Caswell}. However, failure always occur near or below $T_+ = 190 \pm 20^{\circ}$C, a value determined by carefully placing $\approx$ 2 mm drops of distilled water onto a heated, polished, concave aluminum surface until the drops did not break up or fizzle upon contact.

\begin{figure}
    \centering
    \includegraphics[width=3.25 in]{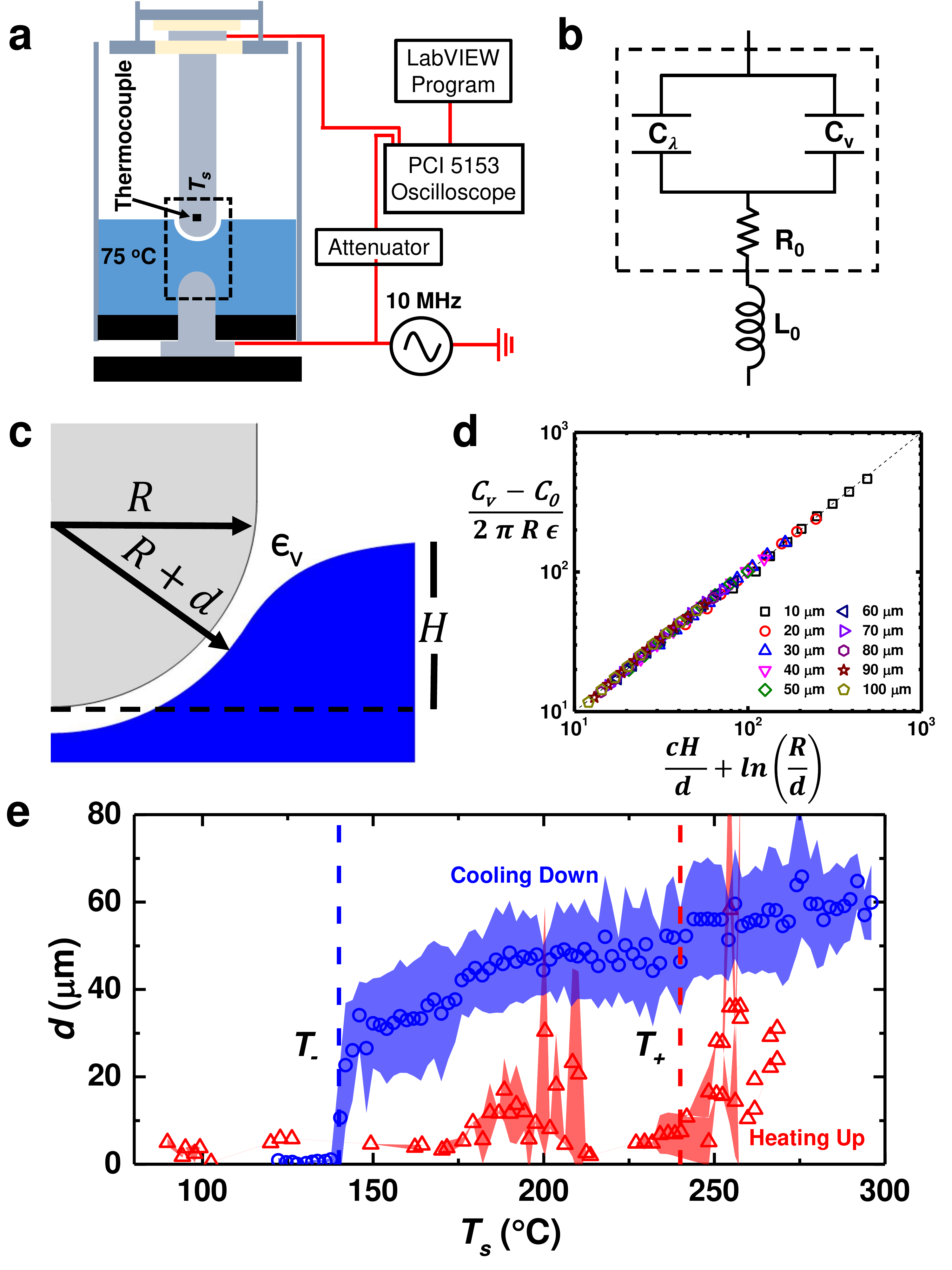}
    \caption{ (a) Experimental setup for high-speed measurements of the vapor layer dynamics. The dashed box indicates the complex impedance which varies in time. (b) Equivalent circuit for the complex impedance of the vapor layer, as described in the main text. Both $C_v$ and $C_\lambda$ are time dependent. (c) Simplified geometry for modeling the capacitance of the vapor layer. (d) Capacitance data from the COMSOL model \cite{supp,comsol} using a range of values for $H$ and $d$. Symbols correspond to different values of $d$. The dashed line is Eq.\ \ref{Cv4}, with $c=0.58$ and $C_0=1.85$ pF. (e) $d$ versus $T_s$ for a nickel-coated copper electrode in pure water. The electrode was first heated (red triangles) and then subsequently cooled (blue circles). $T_+ = 240 \pm 30^{\circ}$C (red dashed line) was the average temperature when the vapor layer formed, and $T_- = 140 \pm 10^{\circ}$C (blue dashed line) was the average temperature when the vapor layer collapsed. The shaded regions show the standard deviation from multiple experiments.}
    \label{Fig2}
\end{figure}

What determines $T_{-}$? Drops on flat surfaces with radii larger than $\approx$ 10 mm are known to succumb to the Rayleigh-Taylor instability \cite{Biance,Snoeijer2009}, yet Fig.\ \ref{Fig1}b shows that all drops fail below a minimum temperature, regardless of size. To better investigate the thickness and dynamics of vapor layers with a well-controlled geometry, we used a heated metallic cylinder (diameter = 7.9 mm) with a rounded tip immersed into a liquid bath heated to a temperature $T_l$ (Fig.\ \ref{Fig2}a) so that a vapor layer forms around it. A ceramic heater and thermocouple were embedded in the cylinder, and the bath was heated externally. With this geometry, both $T_s$ and the water liquid level could be controlled independently.

To study the dynamics of the vapor layer at short time scales, we monitored the electrical impedance between the heated solid and a geometrically-similar lower electrode in the bath. The lower electrode was immersed completely in the liquid, whereas the heated electrode was immersed only to a depth of $H$ (see Fig.\ \ref{Fig2}a). A 10 MHz signal was driven into the lower electrode and then measured at the heated electrode with a PC-based oscilloscope. The amplitude and phase of the signal were extracted by custom software-based lock-in detection (Fig.\ S1 \cite{supp}), as done in similar experiments investigating drop coalescence \cite{Paulsen}. We added a variable concentration of NaCl salt to the bath in order to increase the conductivity of the liquid. 

The region between the two electrodes, shown in the dashed box in Fig.\ \ref{Fig2}a, can be modeled as an RLC circuit (Fig.\ \ref{Fig2}b). The inductance, $L_0$, represents parasitic inductance in the experimental apparatus, and $R_0$ is the combined resistance of the liquid and metal-liquid boundaries. We treat the vapor layer as a capacitor, where one plate is the heated metal electrode and the other is the liquid surface (see Fig.\ \ref{Fig2}c). The interface between the upper, heated electrode and the liquid is modeled as a parallel combination of two capacitors. $C_v$ is the capacitance of the vapor layer, and  $C_\lambda$ is the capacitance of the double layer that forms upon liquid-solid contact (Fig.\ S2 \cite{supp}). When the vapor layer is present, $C_\lambda=0$ and the average vapor layer thickness, $d$, can be computed using a simple geometric model. First, the vapor layer is modeled as a hemispherical capacitor with inner and outer radii equal to $R$ and $R+d$, respectively:

\begin{equation}
    C_1 = 2 \pi \epsilon_{v} \dfrac{R(R+d)}{d}\times c\dfrac{H}{R}.
    \label{Cv1}
\end{equation}
The first term corresponds to a hemispherical capacitor. However, as shown in Fig.\ \ref{Fig2}c, the vapor layer does not encompass the entire volume between the two hemispheres for a given value of the immersion depth $H$. To lowest order, we modified the volume of the vapor layer by a factor of $cH/R$, where $c$ is a numerical constant that takes into account the curved water surface away from the electrode. 

In our experiments, $R = 7.9$ mm and $\epsilon_v=1.0057 \epsilon_0$ is the dielectric constant of water vapor at 100$^{\circ}$C. Since $d/R<1.5\%$, we only consider the leading order term so that
\begin{equation}
    C_1 = 2 \pi c H \epsilon_{v} \dfrac{R}{d}.
    \label{Cv2}
\end{equation}
As the water level is lowered ($H\rightarrow0$), the capacitance approaches that of a sphere above a flat plane. To leading order, this is given by
\begin{equation}
    C_2 = 2 \pi \epsilon_v R \ln\left(\dfrac{R}{d}\right)+C_0.
    \label{Cv3}
\end{equation}
$C_0$ is constant that contains the residual capacitance of the rest of the cylindrical electrode. Combing these terms gives us a relationship between $C_v = C_1+C_2$ and $d$:
\begin{equation}
    \dfrac{C_v-C_0}{2\pi R\epsilon_v} = \dfrac{c H}{d} + \ln\left(\dfrac{R}{d}\right).
    \label{Cv4}
\end{equation}

To verify Eq.\ \ref{Cv4}, we generated the equivalent electrostatic geometry in COMSOL and measured the resulting capacitance, as described in the supplemental material \cite{supp,comsol}. The water surface profile was computed by a surface of revolution composed of two curves: 1) an arc of a circle with radius $R+d$ and 2) a hydrostatic solution of the water's surface profile taking surface tension and gravity into account \cite{Burton2010}. The two curves were matched at a point with continuous first derivatives, providing a unique solution given the boundary conditions (Fig.\ S3 \cite{supp}). By simulating many curves with 1.3 mm $<H<$ 8.3 mm and 10 $\mu$m $<d<$ 100 $\mu$m, we found excellent agreement with Eq.\ \ref{Cv4}, as shown in Fig.\ \ref{Fig2}d. The data collapses well for $c=0.58$ and $C_0=1.85$ pF. To find $d$ from measurements of $C_v$, Eq.\ \ref{Cv4} was analytically solved in terms of product logarithms.  

Upon heating and then cooling the immersed upper electrode, we observed a large, metastable region characterized by hysteresis in $d$ versus $T_s$, in agreement with Fig.\ \ref{Fig1}b. Figure \ref{Fig2}e shows the average vapor layer thickness between pure water at $T_l$ = 75-95$^{\circ}$C and a nickel-coated copper electrode. For this particular experiment, NaCl was not added to avoid salt deposition on the electrode surface. Bubble nucleation and detachment resulted in large variations in our measurements of $d$ during heating. On average, a stable vapor layer formed at $T_+ = 240 \pm 30^{\circ}$C, which is consistent with recent theoretical predictions for metallic surfaces \cite{Zhao}. Nevertheless, once formed, the vapor layer remained stable at temperatures well below $T_+$. As shown in Fig.\ \ref{Fig2}e, the collapse of the vapor layer occurred repeatedly at $T_- = 140 \pm 10^\circ$C. Video S3 \cite{supp} shows a time lapse of a characteristic experiment.   

\begin{figure}[!]
\centering
\includegraphics[width=3.25in]{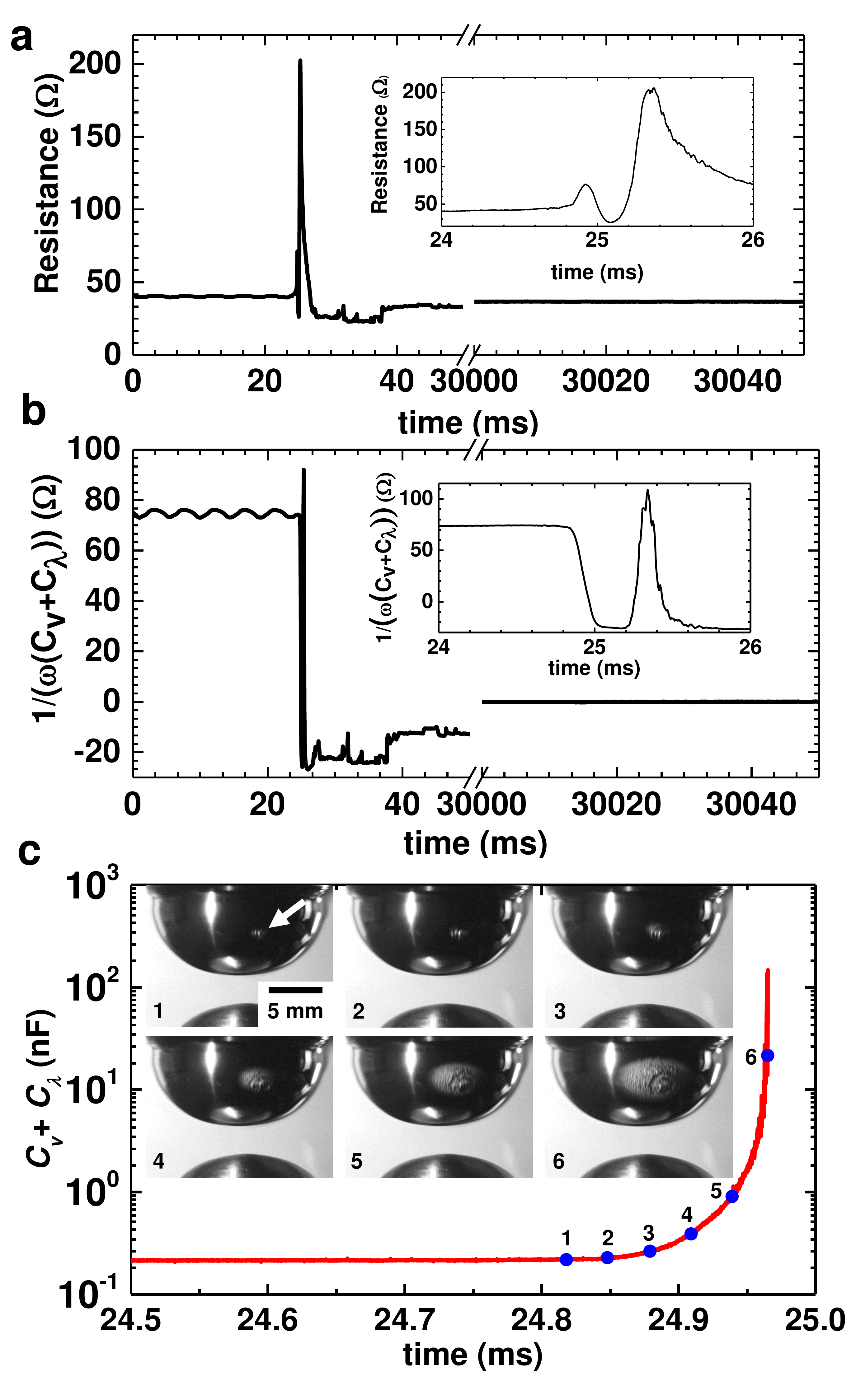}
\caption{Resistance (a) and capacitive reactance (b) of the Leidenfrost cell during collapse with 0.02 M NaCl. The insets show 2 ms of the data right before and after collapse. 30 s after collapse, the system is quiescent. Capillary waves are visible as oscillations in the reactance prior to collapse. (c) Total capacitance of the liquid-vapor-solid interface just before collapse. The enormous increase is due to the formation of an ionic double layer at the liquid-solid contact. The images show a time sequence of the initial collapse point, as indicated by the arrow, where bubbles are generated as the wetting front spreads rapidly. The blue points in the data correspond to the indicated images (Video S4).}
\label{Fig3}
\end{figure}

The collapse of the vapor layer at $T_-$ is explosive and audible. We used synchronized high-speed video (Phantom V7.11, Vision Research) to visualize this process. Figure \ref{Fig3}a and \ref{Fig3}b show the real and imaginary parts of the impedance (dashed box in Fig.\ \ref{Fig2}b) before and after a single collapse event. The bubbles generated during the explosion lead to a large increase in resistance before eventually returning to quiescence at long times. The slightly smaller resistance after collapse was due to the increased water temperature from the heated electrode. For some values of $H$, we observed capillary waves that traveled upwards along the vapor layer. These manifested as oscillations in the reactive impedance (Fig.\ \ref{Fig3}b). In Video S4 \cite{supp}, the capillary waves are visible with a typical wavelength of $\lambda$ = 2-3 mm. We can estimate the corresponding frequency using the dispersion relation for pure capillary waves, $f=(\gamma k^3/\rho_{l})^{1/2}/2\pi\approx$ 120-220 Hz, where $\rho_{l}$ = 959 kg/m$^3$ is the density of pure water at the boiling point, and $k=2\pi/\lambda$ is the wave vector. This agrees well with Fig.\ \ref{Fig3}b, where $f\approx$ 200 Hz. 

During collapse, the combined capacitance, $C_{v}+C_\lambda$, increases by more than 3 orders of magnitude (Fig.\ \ref{Fig3}c), and is facilitated by an explosive wetting front spreading from the initial contact point. The speed of this front is consistent with the capillary velocity, $\gamma/\eta_w\approx 210$ m/s, where $\gamma$ = 59 mN/m is the liquid-vapor surface tension and $\eta_w$ = 0.28 mPa$\cdot$s is the viscosity of water at the boiling point. The large increase in capacitance is due to the formation of an ionic double layer as soon as the liquid contacts the surface, made possible by the addition of salt in the water (Fig.\ S2 \cite{supp}). The effective thickness of the ionic screening layer (1-10 nm \cite{Khademi}) is 3 orders of magnitude smaller than the thickness of the vapor layer, resulting in a much larger capacitance. Thus, even a small fraction of liquid touching the electrode surface will drastically increase the capacitance. The slower decay in the impedance is due to the dissipation of a large cloud of vapor bubbles (Video S4 \cite{supp}). 

\begin{figure}
    \centering
    \includegraphics[width=3.25 in]{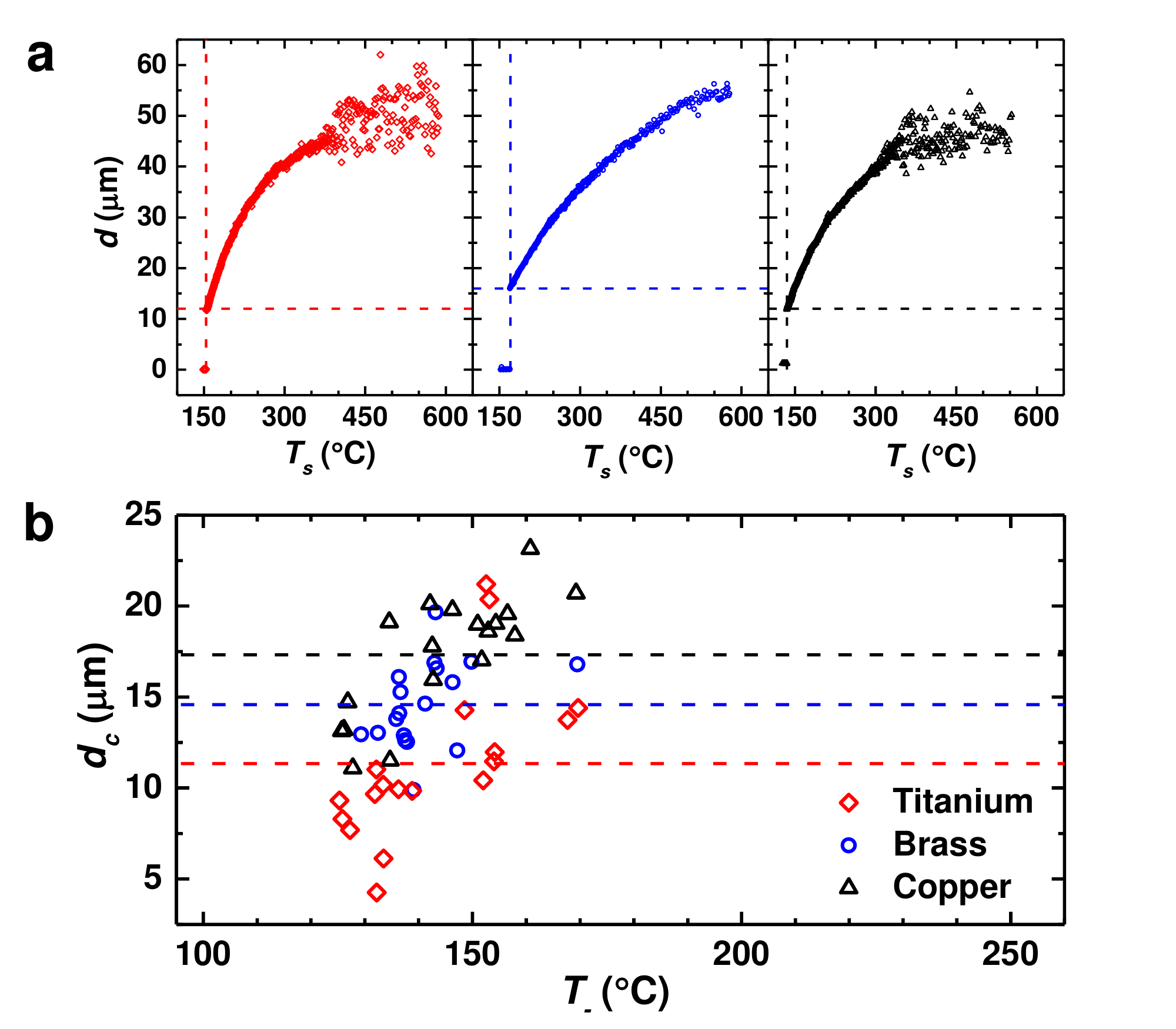}
    \caption{(a) Spatially-averaged vapor layer thickness, $d$, as a function of substrate temperature, $T_{s}$, during the cooling of titanium, brass, and copper electrodes.
    The visible discontinuities in the data, indicated by the dashed lines, correspond to vapor layer failure at temperature $T_-$ and thickness $d_c$ (b) Thickness at failure, $d_c$, as a function of the temperature at failure, $T_-$. Each data point represents varying aqueous NaCl concentrations and liquid level $H$ for each metal. The dashed lines show the mean value of $d_{c}$ for each metal, while the average of $T_-$ was $140\pm 10^{\circ}$C for all experiments.}
    \label{Fig4}
\end{figure}

Surprisingly, and in contrast to the strong dependence of $T_{+}$ on material properties \cite{Zhao}, we found that $T_{-}$ was independent of the metal used for the heated electrode. Figure \ref{Fig4}a shows $d$ versus $T_{s}$ for 3 representative experiments with metals of varying thermal conductivity: titanium (7 W/m$\cdot$K), brass (115 W/m$\cdot$K), and copper (390 W/m$\cdot$K). For each material, the time evolution of $T_{s}$ looked distinct due to differences in heat capacity (Fig.\ S4 \cite{supp}), yet $d$ only depended on $T_{s}$. The discontinuities in the data at lower temperatures mark the failure of the vapor layer and determine both $T_-$ and the spatially-averaged vapor layer thickness at collapse, $d_c$. These values were independently measured in each experiment. Figure \ref{Fig4}b shows $T_-$ and $d_c$ for different metals, liquid levels $H$, and aqueous salt concentrations. A larger thermal conductivity resulted in slightly larger values of $d_c$. We speculate that localized cooling near the liquid interface \cite{Limbeek2017} could result in a smaller vapor layer thickness near the tip of the electrode for metals with lower thermal conductivity. However, the stability of the vapor pocket should be mostly determined by the gas flux through the ``neck" region \cite{Snoeijer2009}, where vapor layer opens up to ambient pressure.

Furthermore, both $T_-$ and $d_c$ were nearly independent over the range 3.7 mm $<$ $H$ $<$ 7.6 mm (Fig.\ S5), which is consistent with the behavior of Leidenfrost drops shown in Fig.\ \ref{Fig1}b. This range of $H$ corresponded to vapor layer surface areas of 89-210 mm$^2$, as computed from the geometric model (Fig.\ S3). Although we did not investigate metal geometries with $R\lesssim l_c$, we would expect a significant drop in $T_-$ in this regime due to a lack of vapor layer fluctuations \cite{Burton,Caswell}. Additionally, $T_-$ and $d_c$ showed no dependence on NaCl salt concentration (Fig.\ S6a). The addition of salt is widely known to suppress Leidenfrost phenomena \cite{Huang,Abdalrahman,Hsu}, despite the fact that NaCl concentrations even up to sea water do not strongly affect the vapor pressure \cite{Fabuss}, evaporation rate, boiling point, viscosity \cite{Kestin}, or surface tension of water \cite{Jones}. Taken together, these measurements suggest that the minimum Leidenfrost temperature is determined by the hydrodynamic stability of the vapor layer. For water, failure reliably occurs at $T_- = 140 \pm 10^\circ$C and $d_c\approx$ 10-20 $\mu$m. 

Although Leidenfrost vapor layers require a material-dependent elevated temperature ($T_+$) for formation, here we showed how vapor layers can be stable at a much lower temperature ($T_-$) that is nearly independent of material and liquid properties. These two temperatures can be separated by more than 100$^{\circ}$C, leading to a large hysteresis and an explosive collapse at low temperatures. This study inherently poses outstanding questions surrounding the initiation of vapor layer collapse, either through unsteady hydrodynamic fluctuations or surface roughness. The liquid interface must approach the surface on sub-micron length scales for short-ranged Van der Waals forces to initiate contact and wetting \cite{Zhao}. We suspect that in highly-dynamic geometries where the vapor layer is constantly in motion, hysteresis may not be visible due to repeated liquid-solid contacts \cite{Jones2019}. Nevertheless, this study explains the surprising robustness of Leidenfrost vapor layers once they are formed, and the physics that determines their violent demise.

\begin{acknowledgments} 
This work was supported by the NSF DMR Grant No. 1455086. 
\end{acknowledgments} 

\end{document}